\documentclass{article}
\usepackage[utf8]{inputenc}
\usepackage[a4paper, total={6.5in, 8.5in}]{geometry}
\usepackage{amsthm}
\usepackage{graphicx} 
\usepackage{array} 
\usepackage{algpseudocode,algorithm,algorithmicx}

\usepackage{amsmath, amssymb, amsfonts, verbatim}
\usepackage{hyphenat,epsfig,subfigure,multirow}
\usepackage{nicefrac}

\usepackage{mathpazo}
\usepackage{enumitem} 
\setlist[enumerate]{wide = 0pt, leftmargin=*}

\usepackage[usenames,dvipsnames]{xcolor}

\usepackage{tcolorbox}
\tcbuselibrary{skins,breakable}
\tcbset{enhanced jigsaw}

\definecolor{@SubtleColor}{rgb}{0,0,.50}

\newcounter{@margincounter}

\newcommand{\ASRQ}{\mathsf{ASRQ}}

\usepackage[normalem]{ulem}
\usepackage[compact]{titlesec}

\definecolor{DarkRed}{rgb}{0.5,0.1,0.1}
\definecolor{RURed}{rgb}{0.95,0.0,0.1}
\definecolor{DarkBlue}{rgb}{0.1,0.1,0.5}

\usepackage{nameref}
\usepackage{mdframed}
\definecolor{ForestGreen}{rgb}{0.1333,0.5451,0.1333}
\definecolor{Red}{rgb}{0.9,0,0}
\usepackage[linktocpage=true,
	pagebackref=true,colorlinks,
	linkcolor=DarkRed,citecolor=ForestGreen,
	bookmarks,bookmarksopen,bookmarksnumbered]
	{hyperref}
\usepackage[noabbrev,nameinlink]{cleveref}
\crefname{property}{property}{Property}
\creflabelformat{property}{(#1)#2#3}
\crefname{equation}{eq}{Eq}
\creflabelformat{equation}{(#1)#2#3}

\newtheorem{mdresult}{Result}

\newtheorem*{mainmdresult}{Main Result}

\usepackage{bm}
\usepackage{url}
\usepackage{xspace}
\usepackage[mathscr]{euscript}
 \usepackage{setspace}
 
\usepackage{tikz}
\usetikzlibrary{arrows}
\usetikzlibrary{arrows.meta}
\usetikzlibrary{shapes}
\usetikzlibrary{backgrounds}
\usetikzlibrary{positioning}
\usetikzlibrary{decorations.markings}
\usetikzlibrary{patterns}
\usetikzlibrary{calc}
\usetikzlibrary{fit}

\newtheorem{theorem}{Theorem}
\newtheorem{lemma}{Lemma}[section]
\newtheorem{proposition}[lemma]{Proposition}

\newtheorem{claim}[lemma]{Claim}

\newtheorem{observation}[lemma]{Observation}

\newtheorem*{claim*}{Claim}
\newtheorem*{proposition*}{Proposition}
\newtheorem*{lemma*}{Lemma}
\newtheorem*{corollary*}{Corollary}
\newtheorem*{remark*}{Remark}

\theoremstyle{definition}
\newtheorem{definition}{Definition}

\newtheorem*{problem*}{Problem}
\newtheorem{remark}{Remark}

\newtheorem{mdalg}{Algorithm}

\newcommand{\eps}{\ensuremath{\varepsilon}}

\newcommand{\bracket}[1]{\left[#1\right]}
\newcommand{\paren}[1]{\ensuremath{\left(#1\right)}\xspace}

\renewcommand{\epsilon}{\varepsilon}
\newcommand{\card}[1]{\left\vert{#1}\right\vert}

\newcommand{\CC}{\ensuremath{\mathcal{C}}}

\newcommand{\norm}[1]{\ensuremath{\left\|#1\right\|}}
\newcommand{\abs}[1]{\ensuremath{\left|#1\right|}}

\newcommand{\prob}[1]{\Pr\bracket{#1}}
\newcommand{\expect}[1]{\Exp\bracket{#1}}

\newcommand{\set}[1]{\ensuremath{\left\{ #1 \right\}}}
\newcommand{\poly}{\mbox{\rm poly}}

\DeclareMathOperator*{\Exp}{\ensuremath{{\mathbb{E}}}}
\DeclareMathOperator*{\Prob}{\ensuremath{\textnormal{Pr}}}
\renewcommand{\Pr}{\Prob}

\newenvironment{tbox}{\begin{tcolorbox}[
		enlarge top by=5pt,
		enlarge bottom by=5pt,
		 breakable,
		 boxsep=0pt,
                  left=4pt,
                  right=4pt,
                  top=10pt,
                  arc=0pt,
                  boxrule=1pt,toprule=1pt,
                  colback=white
                  ]
	}
{\end{tcolorbox}}

\newcommand{\estar}{\ensuremath{e^{\star}}}

\DeclareMathOperator*{\argmin}{argmin}

\newcommand{\Lap}[1]{{\ensuremath{\textnormal{\textsf{Lap}}(#1)}\xspace}}

\newcommand{\dhat}{\ensuremath{\widehat{\dist}}\xspace}

\newcommand{\fhat}{\ensuremath{\widehat{f}}\xspace}

\newcommand{\bx}{\mathbf{x}}

\newcommand{\real}{\mathbb{R}}

\newcommand{\Otilde}{\ensuremath{\widetilde{O}}\xspace}

\newcommand{\dist}{\mathsf{d}}
\newcommand{\vertexset}{\mathcal{V}\xspace}
\newcommand{\edgeset}{\mathcal{E}\xspace}
\newcommand{\graphG}{\mathcal{G}\xspace}

\newcommand{\sampleset}{\mathcal{S}\xspace}

\newcommand{\canonsegset}[1]{\textsf{Canon}(#1)\xspace}

\newcommand{\newepsalgname}{\ensuremath{\textnormal{\textsc{Canon-APSD}}}\xspace}
\newcommand{\approxalg}{\ensuremath{\textnormal{\textsc{SSSP-ASRQ}}}\xspace}

\newcommand{\wtilde}{\ensuremath{\widetilde{w}}\xspace}

\def\S{\mathcal{S}}
\def\R{\mathcal{R}}
\def\reals{\mathbb{R}}
\DeclareMathOperator{\pre}{pred}
\DeclareMathOperator{\suc}{succ}

\usepackage{listings}
\title{Differentially Private Range Query on Shortest Paths}
\author{
Chengyuan Deng \thanks{Rutgers University. email: {\sf cd751@rutgers.edu}.} 
\and 
Jie Gao  \thanks{Rutgers University. email: {\sf jg1555@rutgers.edu}.}
\and 
Jalaj Upadhyay \thanks{Rutgers University. email: {\sf jalaj.upadhyay@rutgers.edu}.}
\and 
Chen Wang \thanks{Rutgers University. email: {\sf wc497@rutgers.edu}.}
}

\begin{document}
\maketitle

\pagenumbering{roman}
\begin{abstract}
    We consider differentially private range queries on a graph where query ranges are defined as the set of edges on a shortest path of the graph. Edges in the graph carry sensitive attributes and the goal is to report the sum of these attributes on a shortest path for counting query or the minimum of the attributes in a bottleneck query. We use differential privacy to ensure that the release of these query answers provide protection of the privacy of the sensitive edge attributes. Our goal is to develop mechanisms that minimize the additive error of the reported answers with the given privacy budget. 
    
    In this paper we report non-trivial results for private range queries on shortest paths. For counting range queries we can achieve an additive error of $\widetilde O(n^{1/3})$ for $\eps$-DP and $\widetilde O(n^{1/4})$ for $(\eps, \delta)$-DP. We present two algorithms where we control the final error by carefully balancing perturbation added to the edge attributes directly versus perturbation added to a subset of range query answers (which can be used for other range queries). 
    Bottleneck range queries are easier and can be answered with polylogarithmic additive errors using standard techniques. 
    
\end{abstract}

\clearpage

\setcounter{tocdepth}{3}
\tableofcontents
\pagebreak

\pagenumbering{arabic}
\setcounter{page}{1}


\section{Introduction}
\label{sec:intro}

Range counting has been extensively studied in the literature, particularly for geometric ranges. In the typical setting, there is a set of points $X$ in $\reals^d$. A range query is often formulated by a geometric shape, and range counting reports the number of points inside the range~\cite{Matousek_1999-hj}. The points can be weighted, in which case the goal is to return the weighted sum inside the query range. 
Compared to the huge literature on geometric range queries~\cite{Toth2017-du}, there has been much less work on the study of range queries with non-geometric ranges.

In this paper, we study private range counting when the ranges are defined as paths on a graph. This setting becomes interesting with the exploding amount of graph data. Graphs are used as a natural mathematical structure to model pairwise relations between objects. Often, the pairwise relations or attributes can represent private and confidential information. 
As such, performing statistics on such a graph without any robust privacy guarantee can be problematic. 
We consider the scenario where both the graph topology and the query ranges (paths on the graph) are public information, but attributes on the edges of the graph, that may come from private sources, are sensitive and protected. Our goal is to return (approximate) range queries while protecting data privacy. 

The above model is applicable in many real-world scenarios. 
In financial analysis, graph-based techniques have been adopted to combat fraud~\cite{Pourhabibi2020-on}.
One can consider a graph where edges represent transactions between two financial entities with attributes such as the total amount being transferred. Forensic analysis researchers may want to issue queries along certain paths that involve multiple financial entities to detect anomalies. 
In supply chain networks, vertices represent participants such as producers, transporters or retailers, and edges represent their relationships. Resilience is a critical factor in supply chains and metrics on edges such as Time-to-Stockout (TTS)~\cite{Hong2022-rh} have been used for estimating end-to-end resilience of certain paths. Response time or cost are also important edge attributes. In these settings, privacy and security issues of the attributes are natural and crucial (e.g., as trade secrets)~\cite{Ogbuke2022-xm}. In road networks, ranges can be naturally defined as paths that users take and queries are about collective statistics of traffic along the path. Privacy is also crucial in healthcare information systems~\cite{Sharma2018-hq}. 

\subsection{Our Setting and Results}

We consider the setting when query ranges are taken as shortest paths based on \emph{public} edge weights, and the query answer is a function of \emph{private} attributes on the edges involved in a query range/path. 
Using shortest paths between two vertices is natural in many of the application settings discussed above. Further, if the range query is applied on arbitrary paths in a graph, the additive query error needed to ensure privacy can be as large as $\Omega(n)$, where $n$ is the number of vertices in the graph. We give a proof of this in \Cref{sec:allpath}.


We consider two types of query function $f$ on a  path $P$:
\begin{itemize}
\item  \emph{Counting query}: return the sum of the attribute values on edges of $P$;
\item \emph{Bottleneck query}: return the minimum of the attribute values on edges of $P$.
\end{itemize} 

Since the attribute values are private and sensitive, the reported range query answers are perturbed to ensure differential privacy guarantees.  Specifically, we consider two neighboring attribute value sets $w$ and $w'$ on the same graph $G$, which differ by a $\ell_1$ norm of $1$. A query mechanism $\mathcal{A}$ is called $(\eps, \delta)$-differentially private if the probability of obtaining query outputs  on input attributes $w$ or $w'$ is relatively bounded by a multiplicative error of $e^{\eps}$ and an additive error of $\delta$.
When $\delta=0$, we call $\mathcal{A}$ $\eps$-DP or pure-DP. The objective is to achieve the specified privacy requirement with noise perturbation as small as possible. 

In this paper, we study the private range query (both counting and bottleneck) on the shortest paths. As standard in the literature of differential privacy, our aim is to understand the trade-off and privacy and additive error in the final query answer, i.e., for a given privacy budget, minimize the additive error. One can additionally consider the query time and space required for the data structure. We leave designing a differentially private data structure with a better query time-space trade-off as a direction of future research. 

For counting queries, we present two algorithms with privacy guarantees of pure-DP and approximate-DP respectively (in \Cref{sec:pure-count-alg} and ~\Cref{sec:alg-approx}), returning the counts with relatively small worst-case additive errors. Our main results are captured by the following theorem:

\begin{mdresult}[$\eps$-DP algorithm for counting query, informal version of \Cref{thm:pure-count-alg}]
\label{thm:pure-count-infm}
There exists an $\eps$-differentially private algorithm that outputs counting queries along all pairs shortest paths with additive error at most $\widetilde{O}(\frac{n^{1/3}}{\eps})$ with high probability.
\end{mdresult}

\begin{mdresult}[$(\eps,\delta)$-DP algorithm for counting query, informal version of \Cref{thm:aprox-count-alg}]
\label{thm:apx-count-infm}
 There exists an $(\eps, \delta)$-differentially private algorithm that outputs counting queries along all pairs shortest paths with additive error at most $\widetilde{O}(\frac{n^{1/4}}{\eps}{\,\, \log^{1/2}\frac{1}{\delta}})$ with high probability.
\end{mdresult}

The above results are the first known upper bounds for this specific problem. Meanwhile, we establish a lower bound of $\Omega(n^{1/6})$ adapted from the construction of the lower bound for private all pairs shortest distances~\cite{ghazi2022differentially} (with details in \Cref{sec:lowerbound}).  The gap between the best-known upper and lower bounds provokes an interesting perspective of private range queries: we do not yet have optimal bounds for specific ranges, despite the results by~\cite{muthukrishnan2012optimal} presenting optimal bounds for generic range query problems. Closing the gap for counting queries would also be an interesting open question.
Our next result, however, shows that the bottleneck query yields simple algorithms using existing techniques to achieve logarithm additive error:

\begin{mdresult}[DP algorithms for bottleneck query, informal version of ]
    \label{thm:apx-bottleneck-infm}
    There exists an $\eps$-differentially private algorithm and an $(\eps, \delta)$-differentially private algorithm, such that with high probability, outputs bottleneck queries along all pairs shortest paths with additive error at most $\widetilde{O}(\frac{\log n}{\eps})$ and $\widetilde{O}(\frac{\sqrt{\log n \, \log \frac{1}{\delta}}}{\eps})$ respectively.
\end{mdresult}

Collectively, our results give the first set of non-trivial bounds for privately releasing queries for shortest paths on range query systems. 
We further show that it is possible to use the VC-dimension of shortest paths queries to obtain a bound similar to \Cref{thm:apx-count-infm}, albeit with a much more complicated algorithm for generic range query applications from \cite{muthukrishnan2012optimal}.

\subsection{Main Techniques}
\label{subsec:tech-overview}
In general, differentially private mechanisms add perturbation to data samples. There are two standard primitives, namely \emph{output perturbation}, where random noises are added to the final data output, and \emph{input perturbation}, where random noises are added to each data element.

We first explain the challenges in improving these two mechanisms. To guarantee privacy, the noise in the output perturbation should take a magnitude of the sensitivity of the range query function. If the edge attribute changes by $1$ in the $\ell_{1}$ norm, there can be up to $\Theta(n^2)$ query pairs being impacted -- e.g., when $\Theta(n^2)$ shortest paths share one edge.
As such, if we apply a crude output perturbation, the noise for each query should be $\widetilde{O}(n^2)$ for $\eps$-differential privacy and $\widetilde{O}(n)$ for $(\eps, \delta)$-differential privacy. On the other hand, with input perturbation, one can add a Laplace noise of magnitude proportional to $1/\eps$ to each edge attribute. This satisfies $\eps$-privacy, but the shortest path may have up to order $n$ edges, and the noises on edges are accumulated with a total error of $\widetilde{O}(n)$.

To improve the error bound, we actually need to combine input and output perturbations. 
In general, the error due to output perturbation is defined by the \emph{sensitivity} of the function -- how many entries will be changed when we have neighboring attributes.
The error for input perturbation depends on the \emph{graph hop diameter}, i.e., the maximum number of edge attributes that we need to sum up as the output of counting queries. 
Therefore, one natural idea is to introduce `shortcuts' (to replace a selective set of shortest paths) to the graph such that the network diameter is reduced. We then apply output perturbation on the shortcuts and use input perturbation on the graph with shortcuts.
Of course, when the shortcuts are introduced, we need to be mindful of their sensitivity.
The natural question is, can we reduce the network diameter with no increase or limited increase to the edge sensitivity with the introduction of the shortcuts? 

\paragraph{Pure-DP algorithm.} The main idea in our first solution is to choose shortcuts with small sensitivity. By the assumption of unique shortest path, any two shortest paths would either be completely disjoint or intersect at \emph{exactly one} common sub-path. 
For every intersecting shortest path between vertices 
$(u_1, u_2) \in \vertexset \times \vertexset$, we name $u_1, u_2$ as the \emph{cut vertices}. 
Since there are $s \choose 2$ shortest path for all pairs in $\sampleset$, there are at most $O(s^2)$ cut vertices on any shortest path $P(u, v)$ with $(u, v) \in \sampleset \times \sampleset$. 
For every $(u,v)\in \sampleset \times \sampleset$, we cut the path $P(u, v)$ along these cut vertices into $O(s^2)$ \emph{canonical segments} and pre-compute their length using output perturbation. 
The good thing is that the maximum sensitivity for the length of a canonical segment is one -- since no two canonical segments can share any common edge. Reducing sensitivity by a multiplicative factor of $s^2$ at the cost of increasing the hop diameter by an additive value of $s^2$ turns out to be beneficial when we calculate the final additive error, which is $\widetilde{O}(\sqrt{n/s+s^2})$,  for our $\epsilon$-DP algorithm. Plugging in $s=n^{1/3}$, we can get an error of $\widetilde{O}(n^{1/3})$ and an $\eps$-DP algorithm. 

\paragraph{Approximate-DP algorithm.} Our solution for $(\eps, \delta)$-DP exploits properties of strong composition~\cite{dwork2010boosting}, which allows us to massage $k$ $(\eps,\delta)$-DP mechanisms into an $(\eps', \delta')$-DP mechanism, where $\eps' \approx \eps \sqrt{k}$ and $\delta' \approx k\delta$.
Our strategy to leverage strong decomposition is to build a shortest path tree rooted at each vertex in the sampled set $\sampleset$. Tree graphs admit much better differentially private mechanisms -- one can get polylogarithmic additive error for running queries on a tree graph~\cite{sealfon2016shortest,fan2022distances}. Now for any two vertices $u, v$ in $\mathcal{G}$, if the shortest path $P(u, v)$ has more than $\Otilde(n/s)$ vertices, $P(u, v)$ has at least one vertex $w$ in $\sampleset$ with high probability. Thus the length of $P(u, v)$ is taken as the sum of length $P(u, w)$ and $P(w, v)$, which, can be obtained by using pre-computed query values between $(u,w)$ and $(v,w)$ in the shortest path tree rooted at $w$. The sensitivity of an edge in this case goes up -- an edge can appear in possibly all the $s$ trees. Thus, on the trees we take $(O(\eps/\sqrt{s}), \delta/2s)$-differentially private mechanisms. The composition of $s$ of them gives $(\eps, \delta)$-DP. The final error bound is $\widetilde O\paren{\sqrt{n/s} +  \sqrt{s} }$. 
Optimizing the error by setting $s=\widetilde O(\sqrt{n})$ gives an $(\eps, \delta)$-DP mechanism with an additive error of $\widetilde{O}(n^{1/4})$.

\begin{remark}
\label{rem:pureDP_tree_based}
Our scheme for the approximate-DP algorithm can also be applied to the pure-DP regime to obtain the same upper bound of $\widetilde{O}({n^{1/3}})$, using the basic composition theorem (\Cref{prop:basic-comp}) and replacing Gaussian mechanism with Laplace mechanism. However, there will be an extra $\log^2 n$ on the additive error over the pure-DP algorithm described above.
\end{remark}

\begin{remark}
The algorithm using canonical segments is only for undirected graphs, while the algorithm using shortest path trees can be extended for directed graphs. In particular, we can build two shortest path trees at each sampled vertex $w$, one $T_{in}(w)$ with edges pointing towards $w$  and one tree $T_{out}(w)$ with edges pointing away from $w$. Any shortest path $P(u, v)$ that visits a vertex $w \in S$ is composed of the shortest path from $u$ to $w$ (captured in the tree $T_{in}(w)$) and then a path from $w$ to $v$ (captured in tree $T_{out}(w)$). With this in mind, throughout the paper we assume an undirected graph.
\end{remark} 

\subsection{Related work}

\paragraph{Geometric Range Queries} Geometric range queries typically consider halfplane ranges, axis-parallel rectangles (orthogonal range query ), or simplices (simplex range query).
The majority of work on range counting considers upper and lower bounds on the running time for answering a query, with different data storage requirements~\cite{Toth2017-du}. 
Designing geometric data structures while preserving differential privacy has also gained attention in the recent past. For example, Biemel et al.~\cite{beimel2019private,kaplan2020private} looked at the problem of the center point of a convex hull. They instantiated exponential mechanism with {\em Tukey depth}~\cite{tukey1975mathematics} as the score function. Since then, several works have looked at various geometric problems, like learning axis-aligned rectangles~\cite{sadigurschi2021sample,beimel2013private}, where one can achieve optimal error bound under pure differential privacy using exponential mechanism; however, the case for approximate differential privacy is still open.   
There has been some recent work that studied differentially private geometric range queries (e.g., orthogonal range queries) under both the {\em central model} and {\em local model} of privacy~\cite{cormode2012differentially,muthukrishnan2012optimal,qardaji2013differentially,cormode2019answering,ghane2021differentially,xiao2010differential,zhang2016privtree}. 

\paragraph{Differentially Private Linear Queries} 
A fundamental class of queries studied in the literature of differential privacy are linear queries on a dataset~\cite{acs2012differentially,bhaskara2012unconditional,blum2005practical,BLR08,bun2018fingerprinting,dwork2006calibrating,gupta2012iterative,hardt2012simple,hardt2010multiplicative,hardt2010geometry,hay2009accurate,hay2009boosting,li2010optimizing,li2013optimal,nikolov2013geometry,qardaji2013understanding,qardaji2014priview,xiao2012dpcube}. Here, given a  dataset from a data universe $\mathcal U$ of size $d$ (usually represented in a form of a histogram $D \in \mathbb R^{d}$) and a query $q \in \mathbb R^d$, the goal is to estimate $q^\top D$. One can replace the query vector with a predicate $\phi:\mathcal U^n \to \set{0,1}$, where $n$ is the size of the database, $D= \set{d_1, \cdots, d_n} \in \mathcal U^n$. The counting query is then simply $\sum_{i=1}^n \phi(d_i)$. Range queries can be seen as a special case of linear queries with a properly defined set of predicates. 

The most relevant work to this paper is the work by Muthukrishnan and  Nikolov~\cite{muthukrishnan2012optimal}, who proposed a differentially private mechanism for answering (generic) range queries when the ranges have bounded VC-dimension~\cite{muthukrishnan2012optimal}. We can apply their techniques to get results for our setting of using shortest paths as ranges. Our algorithm can be easily extended to guarantee $\epsilon$-differentially private with a slight change of parameters, while this substitution is non-trivial for the algorithm of Muthukrishnan and Nikolov~\cite{muthukrishnan2012optimal}, and to the best of our understanding, yields sub-optimal error bound. More discussion of this is in \Cref{sec:vc-dim}.

\paragraph{Private Release of Graph Data}
Private release of graph data has been studied in recent years on many graph properties; see the survey ~\cite{DPgraph-survey}. 
There has been recent work on differentially private release of all pairs shortest path length~\cite{sealfon2016shortest,ghazi2022differentially,fan2022distances,fan2022breaking}. Here, the edge weights $w$ is considered sensitive, and the goal is to produce an approximate distance matrix for all pairs shortest paths length with differential privacy guarantees. 
In other words, the edge weights $w$ \emph{are} the sensitive attributes $a$.
This is a harder problem than the problem 
considered in this paper. Specifically,  the topology of the shortest paths are public information in our setting, but the knowledge of which edges are on the shortest path may reveal knowledge of the sensitive edge length $w$. It has been shown in~\cite{sealfon2016shortest} that when one releases the set of edges on an approximate shortest path in a differentially private manner, the additive error in the distance report has to be as large as $\Omega(n)$.  The best known results for private release of all pairs shortest distance have an additive error of $\widetilde O(n^{2/3})$ for pure-DP and $\widetilde O(\sqrt{n})$ for approximate-DP~\cite{ghazi2022differentially,fan2022distances,fan2022breaking} for general graphs. There is a lower bound of $\Omega(n^{1/6})$ for approximate-DP~\cite{ghazi2022differentially}. 
For trees the two problems are the same since for any two nodes the shortest path is unique regardless of edge length.

Differentially private range query on shortest paths has been done on a planar graph in~\cite{ghosh20differntially}, where they provide mechanisms with polylogarithmic additive error. But this problem has not been studied for the general graph setting.

\section{Preliminaries}

\paragraph{Notation.} 

We use $\mathcal G=(\mathcal V,\mathcal E)$ to denote a graph on vertex set $\mathcal V$ and edges $\mathcal E$. An edge $e\in \mathcal E$ is also denoted by the tuple $(u,v)$ if $u$ and $v$ are its endpoints.  
For a pair of vertices $(u,v)$, we denote $P(u,v)$ as their shortest path, and $d(u,v)$ as the shortest distance. 
We can define the attribute function $w: \edgeset \rightarrow \real^{m}$ over all the edges \emph{independent} of the shortest paths. On a path $P(u,v)$, we let $\gamma(u,v):=\min_{e} \{w(e)| e\in P(u,v)\}$ 
as the minimum attribute value along the shortest path $P(u,v)$.
We use $\R=(X,\S)$ to denote a set system, where $\S$ is a collection of sets with elements from $X$.

\subsection{The Models for Range Query and Privacy}
\paragraph{Shortest Paths as Ranges.} 
In a set system $\R=(X, \S)$, where $X$ is a set of elements, and $\S$ is a collection of subsets $S_i\subseteq X$ called \emph{ranges}. In a graph $\graphG$ when shortest paths are unique\footnote{One can use symbolic perturbation of edge distances to produce unique shortest paths.}, we can define shortest paths as ranges.
We take $X$ to be the set of $m$ edges in $\graphG$, and each set of $\S$ corresponds to a set of edges on a $(u,v)$ shortest path. In particular, for an undirected graph $\graphG$, its corresponding $\S$ has ${n \choose 2}$ order sets; and for a directed graph $\graphG$, $\S$ may have up to $n^2$ ordered sets. 

Based on the set system $\R=(X, \S)$, we can define \emph{range queries} on $\R$ as $(\R, f)$ with a \emph{query function} $f: \S \rightarrow \real$ as $\{f(S)\}_{S \in \S}$ for every set in $\S$. 
We can further extend this notion of range queries on shortest distances with \emph{attribute} functions $w: X \rightarrow \real^{\geq 0}$, and the queries on each set $S$ become $f(w(S))$, where $w(S)$ means to apply attribute function to each element in $S$. 
Note that the attribute function should \emph{not} be considered as edge weights as it does not affect the shortest paths.
Our goal is to release the statistics of \emph{all} sets with small additive errors and privacy guarantees following the definitions in \Cref{def:dp}. 


We now formally define the privacy model for range queries on shortest paths.
\begin{definition}[Range Queries with Neighboring Attributes]
\label{def:neighbor-attributes}
Let $(\R=(X, \S), f)$ be a system of range queries, and let $w, w': X \rightarrow \mathbb{R}^{\geq 0} $ be attribute functions that map each element in $X$ to a non-negative real number. We say the attributes are \emph{neighboring}
\begin{displaymath}
\sum_{x \in X}|w(x)-w'(x)| \leq 1.
\end{displaymath}
We emphasize that the attributes do \emph{not} change the shortest paths, i.e., the graphs operate on the same set system $\R=(X, \S)$. When it is clear from context, we abuse the notation and denote the above by $\norm{w-w'}_1 \leq 1$.
\end{definition}

We shall define the pure- and approximate DP with the notions of the neighboring attributes on range queries as follows.
\begin{definition}[Differentially Private Range Queries]
\label{def:dp}
Let $(\R=(X, \S), f)$ be a system of range queries and $w, w': X \rightarrow \real^{\geq 0} $ be attribute functions as prescribed in \Cref{def:neighbor-attributes}. Furthermore, let $\mathcal{A}$ be an algorithm that takes $(\R, f , w)$ as input. Then $\mathcal{A}$ is $(\varepsilon, \delta)$-differentially private on $\graphG $ if, for all pairs of neighboring attribute functions $w, w'$ and all sets of possible outputs $\CC$, we have that
\begin{displaymath}
\Pr[\mathcal{A}(\R, f , w)\in \CC] \leq e^{\varepsilon}\cdot \Pr[\mathcal{A}(\R, f , w') \in \CC ]+\delta.
\end{displaymath}
If $\delta=0$, we say $\mathcal{A}$ is $\varepsilon$-differentially private on $\graphG $.
\end{definition}

We now define the notion that characterizes the \emph{utility} of the algorithm. In the range query model, we say an algorithm $\mathcal{A}$ provides $(\alpha, \beta)$-approximation to \emph{all sets range queries} ($\ASRQ$) if, given a range query system $(\R=(X, \S), f)$ and a attribute function $w$, with probability at least $1-\beta$, algorithm $\mathcal{A}$  outputs an answer within an $\alpha$ additive error for the original query value on every set.

\begin{definition}[Approximate-$\ASRQ$]
\label{def:priv-rq-utility}

    A randomized algorithm $\mathcal{A}$ is an $(\alpha,\beta)$-approximation for {\em all sets range queries} ($\ASRQ$) on a range query system $(\R=(X, \S), f)$ with attribute function $w$ if for any $S \in \S$, 
    \[
    \prob{ \abs{f(w(S))-\mathcal{A}(w(S))} \leq \alpha} \geq 1 -\beta.
    \]
\end{definition}

Since $S$ contains the ranges of all-pairs shortest paths, the approximation in \Cref{def:priv-rq-utility} naturally corresponds to the additive approximation of shortest distances when $f$ is the \emph{counting query}. Trivially, if we output the range queries simply based on the elements and the attribute function $w$, we have $\alpha=\beta=0$. However, such an output will \emph{not} be private -- and to guarantee both privacy and approximation is the main focus of  this paper. 

\begin{remark}
Our model of \Cref{def:neighbor-attributes} is closely related to the all-pair shortest distances release studied in ~\cite{sealfon2016shortest,ghazi2022differentially,fan2022distances,fan2022breaking}. In particular, in the model of private all-pair shortest distances, the neighboring graphs are also defined as the norm of attributes differing by at most $1$. However, there is a subtle difference: in the shortest distances model, the shortest paths are private and subject to protection; while in the range query model, the shortest paths are known, and we do \emph{not} have to protect their privacy. This allows us to bypass the $\Omega(n)$ additive error lower bound in \cite{sealfon2016shortest} for any algorithm that privately reveal the shortest paths, and obtain much stronger results.
\end{remark}

\subsection{Standard Technical Tools}

\paragraph{Tools from Probability Theory} We first introduce some well-known results from probability theory. We refer interested readers to the standard textbooks on this subject for more details~\cite{wainwright2019high}.

\begin{definition}[Laplace distribution]
\label{def:lap-dist}
We say a zero-mean random variable $X$ follows the Laplace distribution with parameter $b$ (denoted by $X \sim \Lap{b}$) if the probability density function of $X$ follows
\begin{align*}
p(x) = \Lap{b}\paren{x} = \frac{1}{2b}\cdot \exp\paren{-\frac{\card{x}}{b}}.
\end{align*}
\end{definition}

\begin{definition}[Gaussian distribution]
\label{def:gauss-dist}
We say a zero-mean random variable $X$ follows the Gaussian distribution with variance $\sigma^2$ (denoted by $X \sim \mathcal N(0,\sigma^2)$) if the probability density function of $X$ follows
\begin{align*}
p(x) = \frac{1}{\sqrt{2 \pi \sigma^2}}\cdot \exp\paren{-\frac{x^2}{2\sigma^2}}.
\end{align*}
\end{definition}


Both Laplace and Gaussian random variables have nice concentration properties. Furthermore, we can get stronger concentration results by the summation of both random variables~\cite{wainwright2019high}.

\begin{lemma}[Sum of Laplace random variables,~\cite{chan2011private,wainwright2019high}]
\label{lem:lap-sum}
Let $\{X_{i}\}_{i=1}^{m}$ be a collection of independent random variables such that  $X_i \sim \Lap{b_{i}}$ for all $1\leq i \leq m$. Then, for $\nu \geq \sqrt{\sum_{i} b^2_{i}}$ and $0<\lambda<\frac{2\sqrt{2}\nu^2}{b}$ for $b= \max_{i} \set{b_{i}}$, 
\begin{align*}
\prob{\abs{\sum_{i} X_{i}} \geq \lambda}\leq 2\cdot \exp\paren{-\frac{\lambda^2}{8\nu^2}}.
\end{align*}
\end{lemma}

\begin{lemma}[Sum of Gaussian random variables, \cite{wainwright2019high}]
    \label{lem:gas-sum}
Let $\{X_{i}\}_{i=1}^{m}$ be a collection of independent random variables such that  $X_i \sim \mathcal{N}{(\mu, \delta^2)}$ for all $1\leq i \leq m$. Then, 
\begin{align*}
\prob{\abs{\frac{\sum_{i} X_{i} }{m} -\mu} \geq \lambda}\leq 2\cdot \exp\paren{-\frac{m \lambda^2}{2\delta^2}}.
\end{align*}
\end{lemma}


\paragraph{Tools in Differential Privacy}
We proceed to existing tools used frequently in differential privacy:
\begin{definition}[Sensitivity]
\label{def:sensitivity}
Let $p \geq 1$. For any function $f: \mathcal{X}\rightarrow \mathbb{R}^k$ defined over a domain space $\mathcal X$,  the $\ell_p$-sensitivity of the function $f$ is defined as  
\begin{displaymath}
\Delta_{f,p} = \max_{\substack{w,w' \in \mathcal{X} \\
w\sim w'}}\|f(w)-f(w') \|_{p},
\end{displaymath}
Here, $\norm{\bx}_p:=\paren{\sum_{i=1}^d \abs{\bx[i]}^p }^{1/p}$ is the $\ell_p$-norm of the vector $\mathbf x \in \mathbb R^d$ and $\bx[i]$ denote the $i$-th coordinate.
\end{definition}

Based on Laplace distribution, we can now define Laplace mechanism -- a standard DP mechanism that adds noise sampled from Laplace distribution with scale dependent on the $\ell_1$-sensitivity of the function. The formal definition is as follows.

\begin{definition}[Laplace mechanism]
\label{def:lap-mech}
For any function $f: \mathcal{X}\rightarrow \mathbb{R}^k$, the Laplace mechanism on input $w\in \mathcal{X}$ samples $Y_1, \dots, Y_k$ independently from $\Lap{\frac{\Delta_{f,1}}{\varepsilon}}$ and outputs
\begin{displaymath}
M_{\varepsilon}(f) = f(w) + (Y_1, \dots, Y_k).
\end{displaymath}
\end{definition}


The following privacy property of Laplace mechanism is known.
\begin{proposition}[Laplace mechanism~\cite{dwork2006calibrating}]
\label{lem:lap-privacy}
The Laplace mechanism $M_{\varepsilon}(f)$ is $\varepsilon$-differentially private.
\end{proposition}

Similar to Laplace mechanism, we can define the Gaussian mechanism:
\begin{definition}[Gaussian mechanism]
\label{def:gauss-mech}
For any function $f: \mathcal{X}\rightarrow \mathbb{R}^k$, the Gaussian mechanism on input $w\in \mathcal{X}$ samples $Y_1, \dots, Y_k$ independently from $\mathcal N\paren{0,\frac{2\Delta_{f,2}^2 \log(1.25/\delta)}{\varepsilon^2}}$ and outputs
\begin{displaymath}
M_{\varepsilon}(f) = f(w) + (Y_1, \dots, Y_k).
\end{displaymath}
\end{definition}


The following privacy property of Gaussian mechanism is known.
\begin{proposition}[Gaussian mechanism~\cite{dwork2006our}]
\label{lem:gauss-privacy}
For $\epsilon \in (0,1)$, the Gaussian mechanism $M_{\varepsilon;\delta}(f)$ is $(\varepsilon,\delta)$-differentially private.
\end{proposition}

It is well-known that if a mechanism $M$ provides $(\eps,\delta)$-DP output, any function $g$ that takes the output of $M$ as input is also $(\eps,\delta)$-DP. This is known as the \emph{post-processing theorem}, formalized as follows.
\begin{proposition}[Post-processing theorem~\cite{dwork2014algorithmic}]
\label{prop:post-process}
Let $M: \real^{d_{1}} \rightarrow \real^{d_{2}}$ be an $(\eps, \,\delta)$-differentially private mechanism and let $g: \real^{d_2}\rightarrow \real^{d_3}$ be an arbitrary function. Then, the function $g\circ M: \real^{d_1} \rightarrow \real^{d_3}$ is also $(\eps, \,\delta)$-differentially private.
\end{proposition}

Finally, we introduce another useful property of differential privacy:  privacy is preserved when combining multiple differentially private mechanisms even against adaptive adversary. 
\begin{proposition}
[Composition theorem \cite{dwork2006calibrating}]
\label{prop:basic-comp}
For any $\varepsilon > 0$, the adaptive composition of $k$ $\varepsilon$-differentially private algorithms is $k\varepsilon$-differentially private. 
\end{proposition}

\begin{proposition}
[Strong composition theorem \cite{dwork2010boosting}]
\label{prop:strong-comp}
For any $\varepsilon, \delta \geq 0$ and $\delta' > 0$, the adaptive composition of $k$ $(\varepsilon, \delta)$-differentially private algorithms is $(\varepsilon',k\delta+\delta')$-differentially private for
\begin{align*}
\varepsilon' = \sqrt{2k\ln (1/\delta')}\cdot \varepsilon +k\varepsilon(e^\varepsilon-1).
\end{align*}
Furthermore, if $\eps'\in (0,1)$ and $\delta'>0$, the composition of $k$ $\eps$-differentially private mechanism is $(\eps', \delta')$-differentially private for
\begin{align*}
\eps' = \eps\cdot \sqrt{8k\log(\frac{1}{\delta'})}.
\end{align*}
\end{proposition}

The following proposition follows from strong composition theorem. 
\begin{proposition}
[Corollary 3.21 in \cite{dwork2014algorithmic}]
\label{prop:strongDPsmallepsilon}
    Let $\mathcal A_1, \cdots, \mathcal A_k$ be $k$ $(\epsilon',\delta')$-differentially private algorithm for 
    \[
     \epsilon' = \frac{\epsilon}{\sqrt{8k\log(1/\delta)}}.   
    \]
    Then an algorithm $\mathcal A$ formed by adaptive composition of $\mathcal A_1, \cdots, \mathcal A_k$ is $(\epsilon,k\delta'+\delta)$-differentially private.
\end{proposition}



\section{An $\eps$-DP Algorithm for  Counting Queries}
\label{sec:pure-count-alg}

In the current and following section, we focus on private algorithms for the counting query function. As clarified in \Cref{rem:pureDP_tree_based}, the algorithms using single-source shortest-path tree scheme can achieve $\eps$ and $(\eps, \delta)$-DP regime using only different parameters. However, we propose a different algorithmic idea for pure-DP algorithm, which shaves off a $\log^2 n$ factor. We formally state the results on $\eps$-DP as follows.

\begin{theorem}
\label{thm:pure-count-alg}
For any $\eps \geq 0$, there exists an $\eps$-differentially private efficient algorithm that given a graph $\graphG=(\vertexset,\edgeset,w)$ as a range query system $(\R=(X, \S), f, w)$ such that $\S$ is the set of the shortest paths and $f$ is the counting query, with high probability, outputs all pairs counting queries with additive error $O(\frac{n^{1/3}\log^{5/6} n}{\eps})$. That is, the algorithm outputs an estimate $\fhat(\cdot, \cdot)$ such that
\begin{displaymath}
    \Pr\paren{\max_{u,v \in \vertexset}|\fhat(u,v)-f(u,v)| = O\left(\frac{n^{1/3}\log^{5/6} n}{\eps}\right)}\geq 1-\frac{1}{n}.
\end{displaymath}

\end{theorem}

We start with some high-level intuitions. Our algorithm leverages both input-perturbation and output-perturbation, as mentioned in \Cref{subsec:tech-overview}. A naive solution would be applying output-perturbation to the pair-wise counting queries for vertices in $\sampleset$. However, the change of a single edge attribute may trigger the change of potentially every pair of counting queries for vertices in $\sampleset$. As such, by the composition theorem, we need to boost the privacy parameter by a factor of $\card{\sampleset}^2$ since \emph{each} counting query can change by $1$. On the other hand, note that the ranges are shortest paths, which have special structures. With the standard assumption that all shortest paths are unique, two shortest paths only overlap by one common shortest path segment. Therefore instead of using output perturbation directly among vertices in $\sampleset$, we will be better off by decomposing the shortest paths by how they overlap and privatize the decomposed segments. As will become evident, the size of decomposed segments is less than $\card{\sampleset}^2$, hence the cumulative error is reduced.

To formalize the above intuition, we introduce the notion of {\em cut vertices} and {\em canonical segments}. Both notions are defined w.r.t a subset of vertices $\sampleset \subseteq \vertexset$. Informally, a vertex $w$ becomes a cut vertex if it is a vertex of $\sampleset$, or if it witnesses the branching -- either `merging' or `splitting' -- of two shortest paths between different pairs of vertices in $\sampleset$. The formal definition is as follows.

\begin{definition}[Cut Vertices]
\label{def:cut-vertex}
Let $\sampleset\subseteq \vertexset$ be an arbitrary subset of vertices. For any pair of vertices $(u,v) \in \sampleset$ and their shortest path $P(u,v)$, we say $w\in P(u,v)$ is a \emph{cut vertex for $(u,v)$} if it satisfies \emph{one} of the following two conditions:
\begin{enumerate}
\item   $w \in \{u,v\}$; 
\item  $w\not\in \{u,v\}$ and 
\begin{enumerate}
\item $w \in P(x,z)$ for some $(x,z) \in \sampleset$ and $(x,z) \neq (u,v)$;
\item Without any loss of generality,  suppose the path is from $x$. Let $\pre(w)$ be the vertex before $w$ on $P(x,z)$ and  $\suc(w)$ be the vertex after $w$ on $P(x,z)$. Then either $\pre(w) \not\in P(u,v)$ or
$\suc(w) \not\in P(u,v)$.
\end{enumerate}
\end{enumerate}
\end{definition}

\begin{figure}
\centering
\begin{tabular}{cc}
\includegraphics[width=0.4\linewidth]{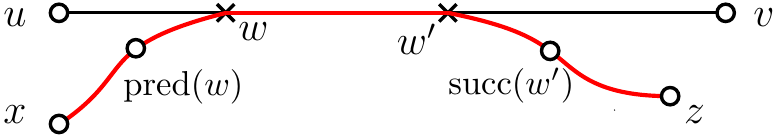} &
\includegraphics[width=0.4\linewidth]{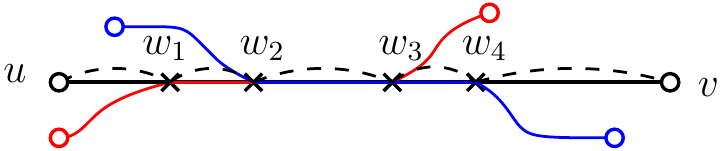}\\
(i) & (ii) \\
\end{tabular}
\caption{(i) Two shortest paths $P(u, v)$ and $P(x, z)$, $u, v, x, z \in \sampleset$, intersect at a common subpath as the shortest path between two cut vertices $w, w'$. (ii) The shortest path $P(u, v)$ is partitioned into canonical segments $P(u, w_1), P(w_1, w_2), \cdots, P(w_{\ell}, w_{\ell+1}), P(w_{\ell+1}, v)$, where $w_1, w_2, \cdots, w_{\ell+1}$ are (ordered) cut vertices along path $P(u, v)$.}
\label{fig:alg-pure}
\end{figure}

See \Cref{fig:alg-pure} (i) for an illustration of cut vertices. Based on \Cref{def:cut-vertex}, we can now define the canonical segments as the path between two adjacent cut vertices along shortest paths of vertices in $\sampleset$.

\begin{definition}[Canonical Segments]
\label{def:can-seg}
Let $\sampleset\subseteq \vertexset$ be an arbitrary subset of vertices. For any pair of vertices $(u,v) \in \sampleset$ and their shortest path $P(u,v)$, a subpath $P(w, w')$ of $P(u, v)$
is a \emph{canonical segment} if 
\begin{enumerate}
\item $w$ is a cut vertex for some $(x,z)\in \sampleset$;
\item $w'$ is a cut vertex for some $(x',z')\in \sampleset$;
\item None of the vertices between $w$ and $w'$ on $P(u, v)$ is a cut vertex for any $(x'',z'')\in \sampleset$.
\end{enumerate}
\end{definition}

Note that $(u,v)$, $(x,z)$, and $(x',z')$ may or may not be the same in the above definition. One can think of the cut vertices as \emph{all} vertices that witnesses the shortest path branching between \emph{all} pairs of vertices in $\sampleset$, and the canonical segments are exactly the collection of segments between adjacent cut vertices along shortest paths of vertices in $\sampleset$. See \Cref{fig:alg-pure} (ii) for an example: $\{u, v, w_{1}, w_{2}, w_{3}, w_{4}\}$ are all cut vertices, which define $5$ canonical segments.

For a fixed vertex pair $(u,v)\in \sampleset$, we define $\canonsegset{\sampleset, u, v}$ as the set of canonical segments on the shortest path of $(u,v)$. Note that the canonical segments need not to be among the edges between the vertices in $\sampleset$:
the shortest path between $(u,v)\in \sampleset$ may well be outside of $\sampleset$. We provide some observations about the basic properties of canonical segments.
\begin{observation}
\label{obs:canon-path-property}
Canonical segments defined as in \Cref{def:can-seg} satisfy the following properties:
\begin{enumerate}
\item\label{line:canon-obs-1} Any two canonical segments are \emph{disjoint}.
\item\label{line:canon-obs-2} The segments in $\canonsegset{\sampleset, u, v}$ covers all edges in $P(u,v)$, i.e. $P(u,v)=\cup_{P(x,z)\in \canonsegset{\sampleset, u, v}}P(x,z)$.
\item\label{line:canon-obs-3} For any pair of vertices $(u,v) \in \sampleset$, there are at most $\card{\sampleset}^{2}$ canonical segments in $\canonsegset{\sampleset, u, v}$ for $\card{\sampleset}\geq 2$.
\end{enumerate}
\end{observation}
\begin{proof}
Observation \ref{line:canon-obs-1} is by definition. Concretely, if two canonical segments overlap, there must be one cut vertex inside another canonical segment, which is not possible by definition.
Observation \ref{line:canon-obs-2} follows from the fact that $u$ and $v$ themselves are cut vertices, and any other cut vertices on $P(u,v)$ only further divides the path. Finally, observation \ref{line:canon-obs-3} holds since every pair of vertices in $\sampleset$ contributes to at
most  two cut vertices on $P(u, v)$. Thus there are at most $2 \cdot {\card{\sampleset} \choose 2}\leq \card{\sampleset}^2$ canonical segments.
\end{proof}

With the definition and properties of canonical segments, we are now ready to present our $\eps$-DP algorithm as follows.

\begin{tbox}
\textbf{\newepsalgname: An $\eps$-DP algorithm to release all pair shortest distances}\\
\textbf{Input: } An $n$ vertices graph, $\graphG = (\vertexset, \edgeset, w)$ and privacy parameter $\varepsilon >0$.
\begin{enumerate}
    \item  Sample a set $\sampleset$ of $s = 100 \zeta \cdot \log n$ vertices uniformly at random, where $\zeta=O(n^{1/3} \log^{-2/3} {n})$
    \item Compute all-pair shortest path for every vertex pair $(x,z) \in \sampleset$ in $\graphG$, and let $P_\sampleset$ be the set of the paths.
    \item Compute $\canonsegset{\sampleset}$ based on the sampled vertices $\sampleset$ and their shortest paths $P_\sampleset$.
    \item \label{line:output-perturb-new-eps-1} \textbf{$\sampleset$ Perturbation:} 
    For each canonical segment $P\in \canonsegset{\sampleset,u,v}$, add an independent Laplace noise $\Lap{2/\eps}$ to its shortest path length.  
    Compute a function $f_{\sampleset}(\cdot, \cdot)$ for counting queries between any vertices $(u,v) \in \sampleset$, by summing up the noisy attributes of the canonical segments in $\canonsegset{\sampleset,u,v}$.
    \item\label{line:input-perturb-new-eps}  \textbf{Non-$\sampleset$ Perturbation:} For each edge in $\graphG$, add independent Laplace noise $\Lap{2/\eps}$ to the edge attribute. 
    For any vertices $u, v\in \vertexset$, let $P(u, v)$ be the shortest path in $\graphG$ and $f'(u,v)$ be the sum of the noisy attributes of the edges along $P(u, v)$. 
    \item For each pair of vertices $(u,v)$,
    \begin{itemize}
        \item If there are at least two vertices in $P(u,v)$ that are in $\sampleset$, let vertex $x$ be the first one along $P(u,v)$ and $z$ be the last one such that $x, z \in \sampleset$, release $\fhat(u,v)=f'(u,x)+f_{\sampleset}(x,z)+f'(z,v)$. 
        \item Otherwise, release $\fhat(u,v) = f'(u,v)$. 
    \end{itemize}
\end{enumerate}
\end{tbox}

We now give the formal analysis of the privacy guarantee and bounds for the additive error.

\subsection{Proof of \Cref{thm:pure-count-alg}}

We start with an observation of the sensitivity of canonical segments. Since canonical segments do not overlap, the weight change of a single edge can only trigger changes of the shortest path distances of at most one canonical segment.
\begin{claim}
\label{clm:low-sensitivity-canon}
Fix any $\sampleset \subseteq \vertexset$, and let $g: (2^{\vertexset}, 2^{\edgeset}) \rightarrow \mathbb{R}^{\card{\canonsegset{\sampleset}}}$ be the function that computes the distances for canonical segments. Then, 
the $\ell_{1}$ sensitivities for $g$ is at most $1$.

\end{claim}
\begin{proof}
The claim follows from the fact that the canonical segments are disjoint (statement 1 of \Cref{obs:canon-path-property}). Concretely, recall that for two neighboring graphs $\graphG \sim \graphG' \in \mathcal{X}$, we have 
$$\sum_{e\in \edgeset}|w(e)-w'(e)|\leq 1.$$ 

As such, 
\begin{align*}
\Delta_{g, 1} &= \max_{\substack{w,w' \in \mathcal{X} \\
w\sim w'}}\norm{g(w)-g(w')}_1 \leq \max_{\substack{w,w' \in \mathcal{X} \\
w\sim w'}}\norm{w-w'}_1 \leq 1,
\end{align*}
where the first inequality follows from the disjointness of canonical segments and the second inequality is by the neighboring graphs.
\end{proof}

Notably, \Cref{clm:low-sensitivity-canon} is already sufficient for us to prove the \emph{privacy} of the algorithm.
\begin{lemma}
\label{lem:eps-alg-new-privacy}
The \newepsalgname algorithm is $\eps$-differentially private.
\end{lemma}

\begin{proof}
We can simply use the (basic) composition theorem (\Cref{prop:basic-comp}) to obtained the desired privacy guarantee. Note that one can view $\sampleset$ Perturbation and Non-$\sampleset$ perturbation as two Laplace mechanisms as defined in \Cref{def:lap-mech}. As such, we only need to prove that both perturbation mechanisms are $O(\eps)$-DP.

By \Cref{clm:low-sensitivity-canon}, the functions in steps~\ref{line:output-perturb-new-eps-1} is of $\ell_{1}$ sensitivity at most $1$. As such, by \Cref{lem:lap-privacy}, its output is $\frac{\eps}{2}$-DP.  For the input perturbation, we are directly operating on the edge attributes. As such, we have $\norm{w-w'}_1\leq 1$. Therefore, by \Cref{lem:lap-privacy}, the $\Lap{\frac{2}{\eps}}$ noise gives an $\eps/2$-DP algorithm. 
\end{proof}

We now proceed to bounding the additive error, which follows a simple idea: we decompose the noise into different parts, and use the concentration of Laplace distribution to get the tight bound.

\begin{lemma}
\label{lem:lap-noise-sum-tail-new}
With high probability, for any vertex pair $(u,v)\in V$, the difference between $f(u,v)$ and $\fhat(u,v)$ released by \newepsalgname is at most $O\paren{\frac{1}{\eps}\cdot \sqrt{\paren{\frac{n}\zeta+\zeta^2\log^2 {n}}\cdot \log n}}$. 
More precisely, 
\begin{align*}
\card{f(u,v) - \fhat(u,v)} 
\leq \frac{900}{\eps}\cdot \sqrt{\paren{\frac{n}{\zeta} +\zeta^2\log^2 {n}}\cdot \log n}
\end{align*}
for any $n \geq C \cdot \zeta \log n$ where $C$ is a sufficiently large absolute constant. 
\end{lemma}
\begin{proof}
We start with proving a structural lemma, which powers the algorithm to decompose the error into different parts to apply the concentration inequality of Laplace noise. The following lemma will be extensively used in the paper:

\begin{lemma}
\label{lem:long-path-divide}
For any pair of vertices $(u,v)$, if the number of edges on the shortest path $P(u,v)$, denoted by $\card{P(u,v)}$, is at least $ \frac{n}{\zeta}$, then, with high probability, there exist at least two vertices $(x, z)\in P(u,v)$ such that
\begin{enumerate}
\item $x\in S$ and $z \in S$.
\item Suppose without any loss of generality, $\card{P(u,x)}\leq \card{P(u,z)}$, then the numbers of edges from $u$ to $x$ and from $z$ to $v$ are at most $\frac{n}{\zeta}$, i.e. $\card{P(u,x)}\leq\frac{n}{\zeta}$ and $\card{P(z,v)}\leq\frac{n}{\zeta}$.
\end{enumerate}
\end{lemma}

We defer the proof \Cref{lem:long-path-divide} to \Cref{sec:proof-long-path}. Now, coming back to the analysis on separate parts of additive error, fix a pair of vertices $(u,v)\in \vertexset$ and their shortest path $P(u,v)$, the additive noises are:
\begin{enumerate}
\item At most $\frac{2n}{\zeta}$ independent noises sampled from $\Lap{\frac{2}{\eps}}$.
\item At most $s^2=100^2 \cdot \zeta^2 \cdot \log^2 {n}$ independent noises sampled from $\Lap{\frac{2}{\eps}}$ for the canonical segments.
\end{enumerate}

The second line is obtained from statements 2 and 3 of \Cref{obs:canon-path-property}: to compute the all-pairs shortest distances between pair in $\sampleset$, it suffices to estimate the canonical segments, and there are at most $s^2$ many of them. As such, in the \newepsalgname algorithm, we let each Laplace noise be with variance $b_i = 2/\epsilon$ for all $i$, 
we again pick $\nu=\sqrt{\sum_{i}b_{i}^2}$ and $\lambda = 30\nu \sqrt{\log n} = \frac{60}{\eps}\cdot \sqrt{n \log n}$. Recall that $s = 100 \log n\cdot n^{1/3}$ (since $\zeta=n^{1/3}$), 
which implies $\frac{2\sqrt{2}\nu}{\max_i b_i}\geq 30\sqrt{\log n}$ (this only needs $n \geq C \cdot \zeta \log n$ for some constant $C$). Therefore, we can apply the concentration of Laplace tail in \Cref{lem:lap-sum}, which gives us
\begin{align*}
\prob{\abs{f(u,v)-\fhat(u,v)}\geq 30\sqrt{\log n} \, \nu}& \leq 2\exp\paren{-\frac{900\log n}{8}} \leq \frac{1}{n^3}.
\end{align*}

Therefore, with probability $1-\frac{1}{n^3}$, 
\begin{align*}
\abs{f(u,v)-\fhat(u,v)} \leq 30\sqrt{\log n}\cdot \nu \leq \frac{90}{\eps}\sqrt{\paren{\frac{n}{\zeta}+100^2\cdot \zeta^2 \cdot \log^2 {n}}\cdot \log n}. 
\end{align*}
A union bound over the above event and the high probability event in \Cref{lem:long-path-divide} gives us the desired statement. 
\end{proof}

In fact, \Cref{lem:lap-noise-sum-tail-new} holds for any $\zeta=n^{1-\Omega(1)}$ for sufficiently large $n$ (as long as $n^{\Omega(1)}>900 \log(n)$). 
We can now finalize the analysis of the additive error of the \newepsalgname algorithm.
\begin{lemma}
\label{lem:eps-alg-new-error}
With high probability, the \newepsalgname algorithm has an additive error of at most $O\paren{\frac{n^{1/3}}{\eps}\cdot \log^{5/6} n}$. 
\end{lemma}
\begin{proof}
We use \Cref{lem:lap-noise-sum-tail-new} by setting the parameter $\zeta=\frac{1}{C} \cdot n^{1/3} \log^{-2/3} {n}$ with the $C$ in \Cref{lem:lap-noise-sum-tail-new}. As such, the total additive error becomes
\begin{align*}
O\paren{\frac{1}{\eps}\cdot \sqrt{\paren{\frac{n}{n^{1/3} \log^{-2/3}{n}}+ (n^{1/3} \log^{-2/3}{n})^{2}\cdot \log^2 {n}}\cdot \log n}} = O\paren{ \frac{n^{1/3}}{\eps}\cdot \log^{5/6} n},
\end{align*}
as claimed. 
\end{proof}

This concludes the proof of \Cref{thm:pure-count-alg}.

\section{A Simple $(\eps, \delta)$-DP Algorithm for Counting Queries}
\label{sec:alg-approx}

Proceeding to the $(\eps,\delta)$-DP setting, we show that with the relaxation of approximate-DP, the worst case additive error can be reduced from $\widetilde{O}(n^{1/3})$ to $\widetilde{O}(n^{1/4})$, formally stated as follows.

\begin{theorem}
\label{thm:aprox-count-alg}
For privacy parameters, $\eps, \delta \in (0,1)$, there exists an $(\eps, \delta)$-differentially private efficient algorithm that given a graph $\graphG=(\vertexset,\edgeset,w)$ as a range query system $(\R=(X, \S), f, w)$ such that $\S$ is the set of the shortest paths and $f$ is the counting query, with high probability, outputs all pairs counting queries with additive error $O\left(\frac{n^{1/4}\log^{2/3} n \log^{1/4} {\frac{1}{\delta}}}{\eps}\right)$. That is, the algorithm outputs an estimate $\fhat(\cdot, \cdot)$ such that
\begin{displaymath}
    \Pr\paren{\max_{u,v \in \vertexset}|\fhat(u,v)-ft(u,v)| = O\left(\frac{n^{1/4}\log^{5/4} n \log^{1/2}{\frac{1}{\delta}}}{\eps}\right)}\geq 1-\frac{1}{n}.
\end{displaymath}

\end{theorem}

 At the high level, our algorithm builds single-source shortest path trees (see formal definition in \Cref{def:sssp}) for each vertex sampled uniformly at random, then employs an $(\varepsilon, \delta)$-DP algorithm for distances release in the tree graph. Notice that the construction of single-source shortest-path trees follows from folklore algorithms based on Dijkstra's algorithm, which takes $O(m+n\log(n))$ time with the classical Fibonacci heap implementation. Further, our algorithm can be easily extended to guarantee $\epsilon$-differentially private with slight change of parameters, while this substitution is non-trivial for the algorithm of Muthukrishnan and Nikolov~\cite{muthukrishnan2012optimal}, and to the best of our understanding, yields suboptimal error bound. \par

\begin{definition}[Single-source shortest-path tree]
\label{def:sssp}
Given a graph $\graphG = (\vertexset, \edgeset)$ and a vertex $s \in \vertexset$, the single-source shortest-path tree rooted at $s$ is a spanning tree $\mathcal{G}'$ such that the unique path from $s$ to $v$ in $\mathcal{G}'$ is the shortest path from $s$ to $v$ in $\graphG$.
\end{definition}

We will use the following result of the $(\varepsilon, \delta)$-DP algorithm for tree graphs (see \Cref{sec:proof-dp-tree}).
\begin{lemma} [$(\epsilon,\delta)$-DP for tree graph]
\label{lemma:dp-tree-alg}
Given a tree graph $\graphG = (\vertexset, \edgeset, w)$ and privacy parameter $\varepsilon,\delta \in (0,1)$, there exists an $(\varepsilon, \delta)$-DP algorithm releasing shortest distances from the root vertex to the rest such that, with high probability, induce additive error at most $O\big(\frac{1}{\varepsilon}\log^{1.5}n\sqrt{\log(\frac{1}{\delta})}\big)$. 
\end{lemma}

We have three remarks for \Cref{lemma:dp-tree-alg}. First for tree graphs, our problem and the private release of all pairs shortest distances are the same -- since there is a unique path between any two vertices in a tree graph. Therefore private release of all pairs shortest distances in a tree graph can be used here. 
Prior work for this problem (\cite{sealfon2016shortest,fan2022distances}) focused on $\epsilon$-DP. Between \cite{sealfon2016shortest,fan2022distances}, Fan and Li's algorithm~\cite{fan2022distances} uses heavy-light decomposition of the tree, with a better error bound only when the tree is shallow. Thus we present the version of $(\epsilon,\delta)$-DP based on Sealfon's algorithm~\cite{sealfon2016shortest}. Second, Seafon's algorithm exploits Laplace mechanism, which is replaced by Gaussian mechanism with $\sigma^2 := 1/\epsilon^2\cdot \ln(1.25/\delta) \log n$ in \Cref{lemma:dp-tree-alg}. Third, the additive error bound for $\epsilon$-DP on tree graph is $O(\frac{1}{\epsilon}\log^{2.5}n)$ with high probability for single-source distance. \Cref{lemma:dp-tree-alg} implies that the $(\epsilon,\delta)$-DP algorithm can shave off a $\log n$ factor, end up with a quadratic improvement  on the logarithm term in the final algorithm for private all pairs shortest distances.

For simplicity, call the algorithm in \Cref{lemma:dp-tree-alg} as $\textsf{PrivateTree}(G)$ with an input tree graph $G$. Also we use $\textsf{SSSP}(v)$ for the single-source shortest path tree algorithm, which takes any $v\in V$ as input and outputs a shortest path tree with $v$ as the root. The $(\varepsilon, \delta)$-DP algorithm is presented above.

\begin{tbox}
\textbf{\approxalg: An $(\eps, \delta)$-DP algorithm to release all pairs counting queries}\\
\textbf{Input: } An $n$ vertices graph, $\graphG = (\vertexset, \edgeset, w)$ and privacy parameter $\varepsilon, \delta >0$.
\begin{enumerate}
    \item Sample a set $\sampleset$ of $s = \zeta \cdot \log n $ vertices uniformly at random, where $ \zeta = O(\sqrt{n} \log^{-2.5}n)$.
    \item For each vertex $v\in \sampleset$, compute $T(v) = \textsf{SSSP}(v)$. Call the set of all trees $\mathsf{T}$.
    \item \textbf{$\sampleset$ Perturbation: }  For each tree $T \in \mathcal{T}$, privatize it by running $ \textsf{PrivateTree}(T)$ with the Gaussian noise $\mathcal{N}\big(\mu = 0, \sigma^2 := \frac{1}{\epsilon_0^2}\ln(1.25/\delta_0) \log n\big)$, $\eps_0, \delta_0$ will be specified later, let the output of count query be $f_{\mathsf{T}}(u,v)$.
    \item \textbf{Non-$\sampleset$ Perturbation: } For each edge in $\graphG$ add independent Gaussian noise $\mathcal{N}\big(\mu = 0, \sigma^2 := \frac{4}{\epsilon^2}\ln(2.5/\delta) \log n\big)$. 
    For any vertices $u, v\in \vertexset$, let $P(u, v)$ be the shortest path in $\graphG$ and $f'(u,v)$ be the sum of the noisy attributes of the edges along $P(u, v)$.
    \item \label{line:approxalg-output}For each pair of vertices $(u,v)$
    \begin{itemize}
        \item If at least one of $u, v$ is in $\sampleset$, release $\fhat(u,v) = f_{\mathsf{T}}(u,v)$.
        \item If $u, v \notin \sampleset$ and the path $P(u,v)$ has one vertex $x \in \sampleset$, release $\fhat(u,v)=f_{\mathsf{T}}(u,x)+f_{\mathsf{T}}(x,v)$.
        \item Otherwise, release $\fhat(u,v) = f'(u,v)$.
    \end{itemize}
\end{enumerate}
\end{tbox}

\subsection{Proof of \Cref{thm:aprox-count-alg}}

Our analysis mainly hinges on the concentration of Laplace random variables (\Cref{lem:lap-sum}), a corollary (\Cref{prop:strongDPsmallepsilon}) of strong composition theorem (\Cref{prop:strong-comp}) and the observation that any shortest path with length larger than $\frac{n}{\zeta}$ goes through at least one vertex in the sampled set $\sampleset$ with high probability (\Cref{lem:long-path-divide}).

\begin{lemma}
\label{lem:sssp-apsd-private}
The \approxalg algorithm is $(\varepsilon, \delta)$-differentially private. 
\end{lemma}
\begin{proof}
First observe that any edge in $\mathcal{G}$ can only appear in at most $s$ trees ($s = \card{\sampleset}$), since we only build one single-source shortest path tree for each vertex in $\sampleset$. Therefore, the $\textsf{PrivateTree}$ algorithm  (\Cref{lemma:dp-tree-alg}) is applied at most $s$ times to any edge. In $\sampleset$ perturbation, the Gaussian mechanism achieves $(\eps_0, \delta_0)$-DP for each tree. Pick $\eps_0, \delta_0$ such that $\eps_0 = \frac{\eps}{4\sqrt{2s\ln(4/\delta)}}$ and $\delta_0 = \frac{\delta}{4s}$, using a corollary of strong composition theorem (\Cref{prop:strongDPsmallepsilon}) on $s$ number of $\textsf{PrivateTree}$ algorithms, we have that the $\sampleset$ perturbation is $(\eps/2, \delta/2)$-differentially private.

Combining with the Non-$\sampleset$ perturbation, which is also $(\eps/2, \delta/2)$-differentially private, it is straightforward to see that the \approxalg algorithm is $(\varepsilon, \delta)$-differentially private.
\end{proof}

The analysis of the additive error is again, similar as in \Cref{thm:pure-count-alg} and \Cref{lem:lap-noise-sum-tail-new}. The only difference is that $s$ takes various values to balance the contribution from output perturbation and the input perturbation, leading to different additive errors.

\begin{lemma}
\label{lem:approx-alg-error}
With high probability, the \approxalg algorithm has additive error at most $O(\frac{n^{1/4}}{\varepsilon}\cdot \log^{1.25}{n}\sqrt{\log {\frac{1}{\delta}}})$
\end{lemma}
\begin{proof}
We first show that with high probability, for any vertex pair $(u,v)\in V$, $\abs{f(u,v)-\fhat(u,v)}$ released by \approxalg is at most $\max \left\{O(\sqrt{\paren{n/\zeta}\log {\frac{1}{\delta}}}/\varepsilon),  O\left(\sqrt{s\log{\frac{2s}{\delta}}}\cdot \log^{1.5}{n}\log \frac{1}{\delta}/\varepsilon\right) \right\}$. 

Notice that the additive error is once again decomposed into noises from `output perturbation' ($\sampleset$ perturbation) and `input perturbation' (Non-$\sampleset$ perturbation). Fix a pair of vertices $(u,v) \in \mathcal{V}$ and  denote their shortest path as $P(u,v)$. By \Cref{lem:long-path-divide} and \Cref{lem:sssp-apsd-private}, the additive noises must be either of the following two cases:
\begin{enumerate}
    \item At most $\frac{2n}{\zeta}$ independent noises sampled from $\mathcal{N}\big(\mu = 0, \sigma^2 := \frac{4}{\epsilon^2}\ln(2.5/\delta) \log n\big)$
   
    \item At most two independent noises induced by the \textsf{PrivateTree} algorithm, which is upper bounded by $O(\frac{2}{\eps_0}\log^{1.5}{n}\sqrt{\log{\frac{1}{\delta_0}}}) $.

\end{enumerate}

The first case considers the third bullet point in Step \ref{line:approxalg-output} of the \approxalg algorithm. From \Cref{lem:long-path-divide}, we know that the additive error is the summation of at most $\frac{2n}{\zeta}$ independent Gaussian noises. The second case considers the first and second points in Step \ref{line:approxalg-output} of the \approxalg algorithm, where $\fhat(u,v)$ is decomposed into two distances output by the \textsf{PrivateTree} algorithm. Notice that only one of the two cases can happen, hence the additive error bound is the maximum of the two. This is different from the analysis in \Cref{lem:lap-noise-sum-tail-new}, where the two cases are combined together to construct the shortest paths. In the following, we give detailed upper bounds of the additive error of two terms.\par 

We now apply the concentration of Gaussian tail (\Cref{lem:gas-sum}) for the first case, 
\begin{align*}
    \prob{\abs{f(u,v)-\fhat(u,v)}\geq t}& \leq 2\exp\paren{-\frac{t^2}{2n/\zeta\cdot \delta^2}},
\end{align*}

Let $t = (n/\zeta)^{1/4}\log^{0.5}n\cdot \delta$, the above probability is smaller than $\frac{1}{n^4}$. Apply union bound on all vertex pairs, then with high probability, then $\abs{\dist(u,v)-\dhat(u,v)}$ for the first case is at most 
\begin{align*}
t = \frac{(n/\zeta)^{1/2}\log^{0.5}n\cdot \delta}{s} = O\left(\frac{1}{\eps}(n/\zeta)^{1/2}\sqrt{\log{\frac{1}{\delta}}}\right)
\end{align*}

Next, we show the additive error in the second case. In the $\sampleset$ perturbation that we pick the privacy parameter $(\eps_0, \delta_0)$ for the Gaussian mechanism where $\eps_0 = \frac{\eps}{4\sqrt{2s\ln(2\delta)}}$ and $\delta_0 = \frac{\delta}{2s}$. 

Recall \Cref{lemma:dp-tree-alg}, the additive error is at most

\begin{align*}
    \frac{1}{\eps_0}\log^{1.5}n\sqrt{\log \frac{1}{\delta_0}} = O\left(\frac{1}{\eps}\sqrt{s\log(\frac{1}{\delta})}\cdot\log^{1.5}n\sqrt{\log{\frac{2s}{\delta}}}\right)
\end{align*}

It only remains to balance the two terms to obtain the maximum additive error. Recall that $s = O(\zeta \cdot \log n)$, we pick $ \zeta = C\sqrt{n} \log^{-2.5}n$, where $C$ is a fixed constant, leading to the following additive error:
\begin{align*}
    O\left(\frac{1}{\eps}\sqrt{\frac{2n}{\zeta} \cdot \log{\frac{1}{\delta}}}\right) = O\left(n^{1/4}\log^{1.25} n \cdot \sqrt{\log{\frac{1}{\delta}}}\right)
\end{align*}

\end{proof}
\newcommand{\lapmin}{\ensuremath{\textsf{\underline{Lap-perturb}}}\xspace}
\newcommand{\gausmin}{\ensuremath{\textsf{\underline{Gaussian-perturb}}}\xspace}
\newcommand{\gammatilde}{\ensuremath{\widetilde{\gamma}}\xspace}

\section{Private algorithms for the Bottleneck Edge Queries}
\label{sec:bottleneck-edge}
We investigate the problem of private \emph{bottleneck edge} queries under the range query model in this section. The problem has natural motivations in a bulk of applications where the resilience on the shortest path is quantified by a \emph{bottleneck} attribute. For instance, in the Time-to-Stockout problem we discussed in \Cref{sec:intro}, the quantity of interest is usually the edge with the \emph{minimum} value of the attribute among the shortest path. We show that we can release such information privately by simply applying the input perturbation technique. More formally, we have:
\begin{theorem}
\label{thm:bottleneck-query}
For privacy parameters $\eps, \delta \in (0,1)$, there exist
\begin{itemize}
    \item an $\eps$-differentially private efficient algorithm that given a graph $\graphG=(\vertexset,\edgeset,w)$ as a range query system $(\R=(X, \S), f, w)$ such that $\S$ is the set of the shortest paths and $f$ is the bottleneck query, with high probability, outputs all pairs bottleneck queries with additive error $O(\frac{\log{n}}{\eps})$. 

    \item an $(\eps, \delta)$-differentially private efficient algorithm that given a graph $\graphG=(\vertexset,\edgeset,w)$ as a range query system $(\R=(X, \S), f, w)$ such that $\S$ is the set of the shortest paths and $f$ is the bottleneck query, with high probability, outputs all pairs bottleneck queries with additive error $O\left(\frac{\sqrt{\log{n}\, \log{\frac{1}{\delta}}}}{\eps}\right)$. 
\end{itemize}
\end{theorem}

\begin{remark}
We remark that the bottleneck edge task cannot be trivially solved by the \emph{top-$k$} selection problem in differential privacy (e.g.~\cite{mcsherry2007mechanism,DBLP:conf/nips/DurfeeR19,QiaoSZ21} and references therein). Note that although it is possible to directly apply top-$1$ selection to privately release the bottleneck edge on \emph{a single} shortest path, the $O(n^2)$-many shortest paths may incur significant privacy loss if we simply use composition. 
\end{remark}

We now present the $\eps$-DP and $(\eps, \delta)$-DP algorithms with the input perturbation technique first developed by \cite{sealfon2016shortest}. Recall that we use $\gamma(u,v)=\min_{e\in P(u,v)}w(e)$ to be the minimum edge weight on the shortest path between $u$ and $v$. Both algorithms can be presented with only differences on a subroutine as follows.
\begin{tbox}
\textbf{Algorithms for minimum attribute edge on the shortest path}

\smallskip

\textbf{Input: } A range system $\R=(X,\S)$ and attribute function $w$, where $X$ and $w$ specifies a graph $\graphG = (\vertexset, \edgeset, w)$, and the ranges $\S$ specifies shortest paths; a privacy budget $\eps, \delta \in (0,1)$.
\begin{enumerate}
\item Perform the input perturbation depending on the application: 
\begin{itemize}
\item For $\eps$-DP, use the $\lapmin$ procedure: add Laplace noise $\Lap{\frac{1}{\eps}}$ to every output of $w$, and obtain $\wtilde$.
\item For $(\eps,\delta)$-DP, use the $\gausmin$ procedure: add Gaussian noise with $\sigma = \frac{\sqrt{2\log(1.25/\delta)}}{\varepsilon}$ to every output of $w$, and obtain $\wtilde$.
\end{itemize}
\item For each shortest path $S\in \S$, find $\estar_{S}$ the edge with the minimum attribute on each shortest path $S\in \S$ with the \emph{original} attribute function $w$, i.e. $\estar_{S} =\argmin_{e} \{w(e)\mid e \in S\}$.
\item Report $\gammatilde = \wtilde(\estar_{S})$ as the attribute of the bottleneck edge on each shortest path $S=P(u,v)$.
\end{enumerate}
\end{tbox}

In other words, the whole algorithm can be framed as adding input noise to the attributes (Laplace noise for $\eps$-DP and Gaussian noise for $(\eps,\delta)$-DP), identifying the bottleneck edge with the \emph{original} attributes, and release the noisy attribute of that bottleneck edge. We now show that the algorithms are differentially private under their respective setting, and the additive error is small.


\subsection*{The Analysis of $\eps$-DP Bottleneck Edge}
The privacy guarantee follows from the input perturbation guarantee and the post-processing theorem (\Cref{prop:post-process}). More formally, we can show the following lemma.
\begin{lemma}
\label{lem:pure-dp-bottleneck-privacy}
The algorithm with $\lapmin$ procedure is $\eps$-differentially private.
\end{lemma}
\begin{proof}
Let $f: \edgeset\rightarrow \real^m$ be the attribute function. By the properties of neighboring attributes (\Cref{def:neighbor-attributes}), it follows that the $\ell_1$ sensitivity $\Delta_{f,1}$ is at most $1$ since the total change of bottleneck edges can be at most $1$. As such, by \Cref{lem:lap-privacy}, the output of $\wtilde$ is $\eps$-DP. Since we only release information as post-processing of $\wtilde$, by \Cref{prop:post-process}, the algorithm is $\eps$-DP.
\end{proof}

We now show that the additive error is bounded by $O(\frac{\log{n}}{\eps})$ with high probability. The argument follows by using the concentration of Laplace distribution and union bound over $\poly(n)$ scenarios.

\begin{lemma}
\label{lem:pure-dp-bottleneck-utility}
If the $\lapmin$ procedure is applied, with high probability, for each pair of vertices $(u,v)$, the difference between the output of $\gammatilde(u,v)$ and the true bottleneck edge attribute $\gamma(u,v)$ is at most $O(\frac{\log{n}}{\eps})$, i.e.
\begin{align*}
    \Pr\paren{\max_{u,v \in \vertexset}|\gamma(u,v)-\gammatilde(u,v)| = O\left(\frac{\log{n}}{\eps}\right)}\geq 1-\frac{1}{n}.
\end{align*}
\end{lemma}
\begin{proof}
For a fixed vertex pair $(u,v)$, we need to take care of at most $n-1$ edges on a shortest path. Note that for each edge $(x,y)$ on the path $P(u,v)$, by the tail bound of Laplace distribution, the error induced by a single Laplace noise is at most $5\frac{\log{n}}{\eps}$ with probability at least $1-\frac{1}{n^5}$. As such, we have
\begin{align*}
\Pr\paren{\max_{e\in P(u,v)} \card{w({e})-\wtilde(e)}>5\cdot \frac{\log{n}}{\eps}}\leq \frac{1}{n^4}.
\end{align*}
Therefore, the additive error on the bottleneck edge is also at most $5\cdot \frac{\log{n}}{\eps}$ with probability at least $1-\frac{1}{n^4}$. Applying a union bound over $n\choose 2$ pairs gives us the desired statement.
\end{proof}

\subsection*{The Analysis of $(\eps, \delta)$-DP Bottleneck Edge}

We now turn to the algorithm for $(\eps,\delta)$-DP. Similar to the case in the $\eps$-DP, we show that the approximate-DP [property holds by the Gaussian noise property and the post-processing theorem. The formal lemma can be shown as follows.
\begin{lemma}
\label{lem:approx-dp-bottleneck-privacy}
The algorithm with $\gausmin$ procedure is $(\eps,\delta)$-differentially private.
\end{lemma}
\begin{proof}
Similar to the proof of \Cref{lem:pure-dp-bottleneck-privacy}, we let attribute function $w: \edgeset\rightarrow \real^m$ be the function of \Cref{def:sensitivity}. We can then bound the $\ell_2$ sensitivity $\Delta_{f,2}$ of the attribute function by $1$, again using the properties of neighboring attributes (\Cref{def:neighbor-attributes}). As such, by \Cref{lem:gauss-privacy} and \Cref{prop:post-process} and the right choice of $\sigma$, the algorithm is $(\eps,\delta)$-DP.
\end{proof}

The benefit of allowing approximate-DP is a quadratic improvement on the additive error -- conceptually, this follows straightforwardly by considering the lighter tail of the Gaussian distribution. We formalize the result as follows.

\begin{lemma}
\label{lem:approx-dp-bottleneck-utility}
If the $\gausmin$ procedure is applied, with high probability, for each pair of vertices $(u,v)$, the difference between the output of $\gammatilde(u,v)$ and the true bottleneck edge attribute $\gamma(u,v)$ is at most $O(\frac{\sqrt{\log{n} \, \log{1/\delta}}}{\eps})$, i.e.
\begin{align*}
    \Pr\paren{\max_{u,v \in \vertexset}|\gamma(u,v)-\gammatilde(u,v)| = O\left(\frac{\sqrt{\log{n} \, \log{\frac{1}{\delta}}}}{\eps}\right)}\geq 1-\frac{1}{n}.
\end{align*}
\end{lemma}
\begin{proof}
Again, for a fixed vertex pair $(u,v)$, there are at most $n-1$ edges among a shortest path. Note that for each edge $(x,y)$ on the path $P(u,v)$, by the tail bound of Gaussian distribution (\Cref{lem:gas-sum}), there is
\begin{align*}
\Pr\paren{\card{\wtilde((x,y))-w((x,y))}>5\sqrt{\log{n}}\sigma}\leq \exp\paren{-10\log{n}}\leq \frac{1}{n^5}.
\end{align*}
As such, with probability at least $1-\frac{1}{n^5}$, the attribute of a single edge is only different from the original with an additive error of $5\sqrt{\log{n}}\sigma$. Therefore, we have
\begin{align*}
\Pr\paren{\max_{e\in P(u,v)} \card{w({e})-\wtilde(e)}>5\sqrt{\log{n}}\sigma}\leq \frac{1}{n^4}.
\end{align*}
By the choice of $\sigma$, we have $5\sqrt{\log{n}}\sigma = O(\frac{\sqrt{\log{n} \, \log{1/\delta}}}{\eps})$. Applying a union bound over $n\choose 2$ pairs gives us the desired statement.
\end{proof}

\section{VC-dimension of Shortest Paths Ranges and Generic Algorithms}
\label{sec:vc-dim}
Under the range query context, it is possible to study the VC-dimension of shortest paths in a graph using a range system. The benefit of such a perspective is that one can apply generic algorithms for private range queries, most notably by the work of Muthukrishnan and Nikolov \cite{muthukrishnan2012optimal}. We discuss the problem from this perspective in this section.

Recall that we say a subset $A\subseteq X$ to be shattered by $\S$ if each of the subsets of $A$ can be obtained as the intersection of some $S\in \S$ with $A$, i.e., if $\S|_A=2^A$. 
The Vapnik–Chervonenkis (VC) $d$ of a set system $(X, \S)$ is defined as the size of the largest subset of $X$ that can be shattered. 
Formally, the definition can be described as follows.
\begin{definition}[Vapnik–Chervonenkis (VC) dimension]
\label{def:vc-dim}
Let $\R=(X,\S)$ be a set system and let $A\subseteq X$ be a set. We say $A$ is shattered by $\S$ if $\{S\cap A \mid S\in \S\} = 2^{A}$, i.e. the union of intersections between sets in $S$ and $A$ covers all subsets of $A$. The Vapnik–Chervonenkis (VC) dimension $d$ of $\R$ is defined as the size of the largest $A\subseteq X$ that can be shattered by $\S$.
\end{definition}

In an undirected graph $G$, the VC-dimension of (unique) shortest paths\footnote{Any set of three vertices $\{u, v, w\}$ cannot be shattered: if one vertex $w$ stays on the shortest path of the other two vertices $u, v$, then one cannot obtain the subset $u, v$; if none of them stays on the shortest path of the other two, then one cannot obtain the subset $u, v, w$ .} is $2$~\cite{Tao2011-qv,Abraham2011-zh}. In a directed graph, the VC-dimension of (unique) shortest paths\footnote{In a directed graph, a directed cycle of $u, v, w$ can be shattered.} is $3$~\cite{Funke2014-sr}.

A closely-related notion is the (primal) shatter function of a set system $\R=(X,\S)$ (with parameter $s$), which is defined as the maximum number of distinct sets in $\{A\cap S \mid S\in\S\}$ for some $A\subseteq X$ such that $\card{A}=S$. More formally, the notion can be defined as follows.
\begin{definition}[Primal Shatter Function]
\label{def:shatter-function}
Let $\R=(X,\S)$ be a set system, and $s$ be a positive integer. The primal shatter function of $\R$, denoted as $\pi_{\R}(s)$, is defined as $\max_{A: \, \card{A}=s}\card{\{A\cap S \mid S\in\S\}}$
\end{definition}
It is well known that if the VC-dimension of a range space is $d$, then $\pi_{\R}(s)=O(s^d)$~\cite{Matousek_1999-hj}. This immediately gives a bound of $O(s^2)$ for shortest paths in undirected graphs. We now show that shortest paths in directed graphs enjoys the same bound as well despite having a higher VC-dimension.

\begin{lemma}
\label{lem:sp-shatter-func}
For a range query system $\R=(X,\S)$ defined by shortest paths in (both directed and undirected) graphs, the primal shatter function is $\pi_{\R}(s)=O(s^2)$ for any $s$. 
\end{lemma}
\begin{proof}
Take any set $A$ of size $s$, any shortest path either does not contain any vertex in $A$, or contains a first vertex $x\in A$ and the last vertex $y\in A$ along the path. Notice that $x, y$ might be the same vertex. Thus $\S|_A$ contains the subset of $A$ as $A\cup S(x, y)$, $\forall x, y\in A$ where $S(x, y)$ is the set of vertices on the shortest path from $x$ to $y$. Therefore $\S|_A$ has at most $O(s^2)$ elements.
\end{proof}


The benefit of understanding the VC-dimension and the primal shatter function for shortest system is that we can use generic algorithms for private range queries. In particular, Muthukrishnan and Nikolov~\cite{muthukrishnan2012optimal} have developed a differentially private mechanism for answering range queries of bounded VC-dimension. The guarantee of the algorithm is as follows.
\begin{proposition}[Muthukrishnan-Nikolov algorithm~\cite{muthukrishnan2012optimal}, rephrased]
\label{prop:MN-generic-alg}
Let $(\R = (X, \S), f)$ be a range query system, where $f$ is the counting query and the primal shatter function of $\R$ is $\pi_{\R}(s)=O(s^d)$ for any $s$. There exists an algorithm that outputs \emph{all queries} with
\begin{itemize}
    \item Expected average squared error of $O\left(\frac{n^{1-1/d}\log{\frac{1}{\delta}}}{\eps^2}\right)$; 
    
    \item With probability at least $1-\beta$, worst case squared error of $O\left(\frac{n^{1-1/d}\log{\frac{1}{\delta}}\log{\frac{n}{\beta}}}{\eps^2}\right)$.
\end{itemize}
The algorithm is $(\eps,\delta)$-differentially private.
\end{proposition}

Using the algorithm of \Cref{prop:MN-generic-alg}, the bound on the primal shatter functions of \Cref{lem:sp-shatter-func}, and the fact that counting query sums up the attributes on the shortest paths, we can obtain an $(\eps,\delta)$-DP result with additive error $O(n^{1/4}\log^{1/2}{(n)}\log^{1/2}{(1/\delta)}/\epsilon)$ with high constant probability. This matches our $(\eps,\delta)$-DP result in \Cref{thm:aprox-count-alg} up to lower order terms.

\begin{remark}
\label{rmk:merit-over-blackbox}
Although it is possible to recover the bound of \Cref{thm:aprox-count-alg} using \Cref{prop:MN-generic-alg} as a black-box, our constructions still enjoy multiple advantages. In particular, the construction of \Cref{prop:MN-generic-alg} does \emph{not} give any non-trivial bound for $\eps$-DP, and it is not trivial to adapt it to pure-DP within the framework. Furthermore, the algorithm of \Cref{prop:MN-generic-alg} requires to find a maximal set of ranges with the minimum symmetric differences on different levels, and by the packing lemma bound in \cite{muthukrishnan2012optimal}, it appears that a straightforward implementation could take $\Theta(n^4)$ time in the worst case. On the other hand, our constructions for both \Cref{thm:pure-count-alg} and \Cref{thm:aprox-count-alg} can be implemented in $\Otilde(n^2)$ time. Finally, the algorithm of \Cref{prop:MN-generic-alg} is much more complicated and counter-intuitive, and our algorithm enjoys much better simplicity. 
\end{remark}

\section{Conclusion and Future Work}
We study the private release of shortest path queries under the range query context in this paper, where the graph topology and the shortest paths are public, and the attributes on the graphs (which do \emph{not} affect shortest paths) are subject to privacy protection. 
Our upper bounds cannot be applied to the (harder) problem of private release of all pairs shortest distances~\cite{sealfon2016shortest}. Thus improving the bounds of private range query problem (with upper bound $\widetilde O(n^{1/3})$ for $\eps$-DP and $\widetilde O(n^{1/4})$ for $(\eps, \delta)$-DP) and all pairs shortest distances release (with upper bound $\widetilde O(n^{2/3})$ for $\eps$-DP and $\widetilde O(n^{1/2})$ for $(\eps, \delta)$-DP), where both have a lower bound of $\Omega(n^{1/6})$, remains an interesting open problem. 
Furthermore, since our algorithms are simple to implement, the empirical performances of our algorithms could be another future research direction.

\subsection*{Acknowledgements} We would like to thank Adam Sealfon, Shyam Narayanan, Justin Chen,  Badih Ghazi, Ravi Kumar, Pasin Manurangsi, Jelani Nelson and Yinzhan Xu for useful discussion and suggestions. This research is supported by Decanal Research Grant.

\bibliographystyle{alpha}
\bibliography{reference, privacy}
\clearpage
\appendix

\section{Range Query on All Paths}\label{sec:allpath}

When we allow queries along any path in a graph and require differential privacy guarantees, the following result provides a lower bound of $\Omega(n)$ on the additive error. To show the lower bound, we first consider a range query formulated by the incidence matrix $A$, with $m$ columns corresponding to the $m$ edges in the graph $G$ and rows corresponding to all queries. A query along path $P$ is represented by a row in the matrix with an element of $1$ corresponding to edge $e$ if $e$ is on $P$ and $0$ otherwise. 
We will then talk about the discrepancy of matrix $A$.

The classical notion of discrepancy of a matrix $A$ is the minimum value of $||Ax||_{\infty}$, where $x$ is a vector with elements taking values $+1$ or $-1$. And the hereditary discrepancy of $A$ is the maximum discrepancy of $A$ limited on any subset of columns. As shown in~\cite{muthukrishnan2012optimal}, both discrepancy and hereditary discrepancy of $A$ provides a lower bound on the additive error of differentially private range query using incidence matrix $A$.



\begin{theorem}
A $(\eps, \delta)$-differential privacy mechanism that answers range queries where ranges are defined on any path of an input graph has to incur additive error of $\Omega(n)$.
\end{theorem}
\begin{proof}
Consider a graph of $n+1$ vertices $v_1, v_2, \cdots, v_{n+1}$ and $2n$ edges. Between vertices $v_i$ and $v_{i+1}$ there are two parallel edges $e_i$ and $e'_i$. On this graph there are $2^n$ paths from $v_1$ to $v_{n+1}$. We consider only queries along these paths and the incidence matrix is a tall matrix $A$ of $2^n$ rows and $2n$ columns, corresponding to the $2n$ edges in the graph. 
Now we take a submatrix of $A$
with only the columns corresponding to edges $e_i$. This gives a matrix $A'$ of $2^n \times n$, with the rows corresponding to all subsets of $[n]$.
$A'$ has discrepancy of $\Omega(n)$. To see that, consider the specific vector $x$ that minimizes $||A'x||_{\infty}$. Suppose $x$ has $k$ entries of $+1$ and $n-k$ entries of $-1$. Without loss of generality, we assume $k\geq n/2$, The row of $A$ that has value $1$ corresponding to the positive entries of $x$ and value $0$ corresponding to the negative entries of $x$, gives a value of $k\geq n/2$. Thus $||Ax||_{\infty}$ is at least $n/2$. This means that the hereditary discrepancy of $A$ is at least $\Omega(n)$.

By the same argument and use Corollary 1 in~\cite{muthukrishnan2012optimal}, we conclude that any $(\eps, \delta)$-differentially private mechanism has to have error of $\Omega(n)$.  
\end{proof}

\section{Proof of \Cref{lemma:dp-tree-alg} -- {$(\eps, \delta)$ Algorithm for Tree Graphs}} 
\label{sec:proof-dp-tree}
\begin{proof}
We first claim that we can answer all pairs shortest distance on a tree with $(\alpha,\beta)$-accuracy for 
\[
\alpha = O\paren{\frac{1}{\epsilon}\log n \sqrt{\log\paren{\frac{n}{\beta}}\log\paren{\frac{1}{\delta}} }} 
\]
showing the utility guarantee of \Cref{lemma:dp-tree-alg}. 
Specifically, if we wish to have high probability bounds for the shortest path distance errors, i.e., $\beta=O(1/n)$, the error is upper bounded by $O\paren{\frac{1}{\epsilon}\log^{1.5} n \sqrt{\log\paren{\frac{1}{\delta}} }}$.


In Sealfon's algorithm~\cite{sealfon2016shortest}, a tree rooted at $v_0$ is partitioned into subtrees each of at most $n/2$ vertices. Specifically, define $v^*$ to be the vertex with at least $n/2$ descendants but none of $v^*$'s children has more than $n/2$ descendants. The tree is partitioned into the subtrees rooted at the children of $v^*$, and a subtree of the remaining vertices rooted at $v_0$. In Sealfon's algorithm a Laplace noise of $\Lap{\log n/\epsilon}$ is added to the shortest path distance from $v_0$ to $v^*$ and the edges from $v^*$ to each of its children. The algorithm then repeatedly privatizes each of the subtrees recursively. 
Using Sealfon's algorithm, we know that for a given root node $v_0$, computing the single source (with the root being the source) shortest path distance requires adding at most $O(\log n)$ privatized edges. Further, their algorithm ensures that any edge can be in at most $\log n$ levels of recursion and hence can be used to compute $O(\log n)$ noisy answers. In other words, the number of adaptive compositions we need is $O(\log n)$. 

We use the Gaussian mechanism to privatize the edges. Since we are concerned with approximate-DP guarantee, the variance of the noise required to preserve $(\epsilon,\delta)$-differential privacy is $\sigma^2 := O\paren{\frac{1}{\epsilon^2}\log(1/\delta) \log n}$. 

Fix a node $u$. Let $\widehat d(u,v_0)$ be the distance estimated by using Sealfon's algorithm instantiated with the Gaussian mechanism instead of the Laplace mechanism. Now the noise added are zero mean. Therefore, 
\[
\mathbb{E} [\widehat d(u,v_0)] = d(u,v_0).
\]
Using the standard concentration of Gaussian distribution~\cite{wainwright2019high} implies that 
\begin{align*}
\mathsf{Pr} \left( \left\vert \widehat d(u,v_0) - 
\mathbb{E} \left[\widehat d(u,v_0) \right] \right \vert > a \right) &\leq 2 e^{-a^2/(2 \sigma^2 \log n)}.
\end{align*}
Setting $a = \frac{C}{\epsilon}\log n \sqrt{\log\paren{\frac{2n}{\beta}}\log\paren{\frac{1}{\delta}} }$ for some constant $C>0$, we have 
\begin{align*}
\mathsf{Pr} \left( \left\vert \widehat d(u,v_0) - 
\mathbb{E} \left[\widehat d(u,v_0) \right] \right \vert > \frac{C}{\epsilon}\log n \sqrt{\log\paren{\frac{2n}{\beta}}\log\paren{\frac{1}{\delta}} } \right) &\leq 2 e^{-C \log(2n/\beta)} \leq \frac{\beta}{ n}.
\end{align*}

Now union bound gives that 
\[
\mathsf{Pr} \left( \max_{u \in \vertexset} \left\vert \widehat d(u,v_0) - 
d(u,v_0)  \right \vert \leq  \frac{C}{\epsilon}\log n \sqrt{\log\paren{\frac{n}{\beta}}\log\paren{\frac{1}{\delta}} } \right) \geq 1- \beta.
\]

We can now use the above result to answer all pair shortest paths by fixing a node $v^*$ to be the root note and compute a single source shortest distance with the root node being the source node. Once we have all these estimates, to compute all pair shortest distance, for any two vertices, $(u,v) \in \vertexset \times \vertexset$, we first compute the least common ancestor $z$ of $u$ and $v$. We then compute the distance as follows:
\[
\widehat d(u,v) = \widehat d(u,v^*) + \widehat d(v,v*) - 2 \widehat d(z,v^*).
\]

Since each of these estimates can be computed with an absolute error $O\paren{\frac{1}{\epsilon}\log n \sqrt{\log\paren{\frac{n}{\beta}}\log\paren{\frac{1}{\delta}} }}$, we get the final additive error bound. That is, 
\[
\Pr\paren{\max_{u,v \in \vertexset}|\dhat(u,v)-\dist(u,v)| = O\paren{\frac{1}{\epsilon}\log n \sqrt{\log\paren{\frac{n}{\beta}}\log\paren{\frac{1}{\delta}} }} }\geq 1-\beta
\]   
completing the proof of the claim. 
\end{proof}

\section{Proof of \Cref{lem:long-path-divide}}
\label{sec:proof-long-path}

\begin{proof}[Proof of \Cref{lem:long-path-divide}]
The lemma is proved by a simple application of the Chernoff bound. For each path $P(u,v)$ with more than $\frac{n}{\zeta}$ edges, let $v'$ be the $\paren{\frac{n}{\zeta}+1}$-th vertices on the path $P(u,v)$ from $u$. Similarly, let $u'$ be the $\paren{\frac{n}{\zeta}+1}$-th vertices on the path $P(u,v)$ from $v$ (traversing backward). We show that there must be two vertices sampled in $S$ on both $P(u, v')$ and $P(u',v)$, which is sufficient to prove the lemma statement.

Define $X_{u,v'}$ as the random variable for the number of vertices on $P(u,v')$ that are sampled in $S$, and define $X_{z}$ for each $z \in P(u,v')$ as the indicator random variable for $z$ to be sampled in $S$. It is straightforward to see that $X_{u,v'}=\sum_{z\in P(u,v')} X_{z}$. Since $P(u,v')$ has at least $\frac{n}{\zeta}$ vertices, and we are sampling $s=100\, \log{n}\cdot \zeta$ vertices uniformly at random as $S$, the expected number of vertices on $P(u,v')$ that are sampled is at least $100\, \log{n}$. Formally, we have
\begin{align*}
\expect{X_{u,v'}}\geq 100\, \log{n}\cdot \frac{\zeta}{n}\cdot \frac{n}{\zeta} = 100\, \log{n}.
\end{align*}
As such, by applying the multiplicative Chernoff bound, we have
\begin{align*}
\prob{X_{u,v'}\leq 2} &\leq \exp\paren{-\frac{0.8^2\cdot 100\, \log{n}}{3}}\\
&\leq \frac{1}{n^{10}}.
\end{align*}
The same argument can be applied to $P(u',v)$ by defining $X_{u',v}$ as the total number of vertices that are sampled in $S$. We omit the repetitive details for simplicity. Finally, although the random variables for different $(u,v)$ pairs are dependent, we can still apply a union bound regardless the dependence, and get the desired statement.
\end{proof}

\section{A Remark on Range Query Shortest Path Lower Bound}\label{sec:lowerbound}

For counting range queries with $(\eps, \delta)$-DP guarantee, there is a lower bound of $\Omega(n^{1/6})$ on the additive error, adapted from the construction of the lower bound for private all pairs shortest distances~\cite{ghazi2022differentially}. Specifically, the construction uses a graph where vertices are points in the plane and edges map to line segments between two points that do not contain other vertices. The edge length is the Euclidean length and therefore the shortest path between two vertices is the path corresponding to a straight line. The range query problem can be now formulated as a (special case) of linear queries, as in Section~\ref{sec:vc-dim} and Section~\ref{sec:allpath}, where the matrix $A$ corresponds to the incidence matrix of the shortest paths and the edges in the graph. It is known that this matrix has a discrepancy lower bound of $\Omega(n^{1/6})$~\cite{Matousek_1999-hj}. By the connection of the discrepancy and linear query lower bounds~\cite{muthukrishnan2012optimal}, this is a lower bound for our problem.


\end{document}